\documentclass[aps,prc,preprint,showpacs]{revtex4-1}
\usepackage{graphicx}
\usepackage[english]{babel}
\usepackage[T1]{fontenc}
\newcommand {\nc} {\newcommand}
\nc {\beq} {\begin{eqnarray}}
\nc {\eeqn} [1] {\label{#1} \end{eqnarray}}
\nc {\eoln} [1] {\label{#1} \\}
\nc {\eol} {\nonumber \\}
\nc {\rref} [1] {(\ref{#1})}
\nc {\Eq} [1] {Eq.~(\ref{#1})}
\nc {\Ref} [1] {Ref.~\cite{#1}}
\nc {\la} {\mbox{$\langle$}}
\nc {\ra} {\mbox{$\rangle$}}
\nc {\Kmax} {\mbox{$K_{\rm max}$}}
\nc {\pb} {\bar{p}}
\nc {\ve} [1] {\mbox{\boldmath $#1$}}
\begin{document}
\title{Quasibound states of an antiproton and a hydrogen atom}
\author{Daniel Baye}
\email[]{dbaye@ulb.ac.be}
\author{J\'er\'emy Dohet-Eraly}
\email[]{jdoheter@ulb.ac.be}
\affiliation{Physique Quantique, and\\
Physique Nucl\'eaire Th\'eorique et Physique Math\'ematique, C.P. 229, Universit\'e libre de Bruxelles (ULB), B-1050 Brussels, Belgium.}
\date{\today}
\begin{abstract}
Accurate three-body quantal calculations of the system composed of a proton, an antiproton, and an electron 
are performed in perimetric coordinates with the Lagrange-mesh method, 
an approximate variational calculation with the simplicity of a calculation on a grid. 
Quasibound states with respect to the $\bar{p}$ + H($n = 2$) threshold are obtained 
for $L = 60$ to 71 for various vibrational excitations. 
Their energies have accuracies up to about $10^{-14}$ atomic units 
while less precise energies are determined for $L = 56-59$ broader resonances. 
Their structure is analyzed with the help of mean distances between the particles. 
These mean distances indicate that the proton-electron subsystem is in excited states, mostly $n = 2$, 
as predicted by Sakimoto (Phys. Rev. A 98, 042503, 2018) with the Born-Oppenheimer approximation. 
A comparison performed with this approximation provides the accuracies of its energies 
and of its proton-antiproton mean distances. 
\end{abstract}
\maketitle
\section{Introduction}
The existence of low-energy antiproton beams allows the start of the realization of a long-awaited goal, 
the study of antimatter. 
Of particular interest is the simplest system, antihydrogen, which is now available in experiments \cite{AAB18}. 
It will allow testing a number of basic properties of antimatter \cite{BBC15}. 
Other systems involving antiprotons are very informative such as the hydrogenlike protonium \cite{De60,ZAA06} 
or antiprotonic atoms where an antiproton replaces an electron. 
Among these atoms, antiprotonic helium composed of a helium nucleus, an antiproton, and an electron 
has been studied in very accurate experiments \cite{YMH02,HDE06,HAS16} 
and theoretical calculations \cite{KBM99,KKK03,KHK14,KHK15,BDS19}. 
Together, they provided a very precise measurement of the antiproton mass. 
Within the present experimental and theoretical error bars, this mass is found equal to the proton mass. 

Antiprotonic helium presents a variation of structures with increasing angular momenta: 
successively hydrogenlike atom, quasistable pseudomolecule, and quasistable Rydberg pseudoatom \cite{BDS19}. 
An interesting question is what happens in such a system where the helium nucleus is replaced by a proton, 
i.e., the antiprotonic H$^-$ ion.  
Because of the smaller charge of the proton, the existence of this antiproton-hydrogen system is far from obvious. 
In a study based on the Born-Oppenheimer approximation, Sakimoto has deduced that the binding 
of an antiproton by an hydrogen atom in its ground state is unlikely \cite{Sa14}. 
In a further work, however, he has shown that this binding could be possible at very high angular momenta 
when the hydrogen atom is in its first excited Stark-like state \cite{Sa18}. 
The reason is that the antiproton-hydrogen Born-Oppenheimer potential then has a much slower 
decrease than for hydrogen in its ground state and can be attractive enough at large distances. 
In a careful study of the decay processes, Sakimoto has shown that long-lived states 
could exist for total orbital momenta between 60 and 73. 

The aim of the present paper is to establish the existence of these quasistable states of 
the proton-antiproton-electron system in a fully quantal three-body calculation. 
To this end, we employ the Lagrange-mesh method \cite{BH86,VMB93,BHV02,HB99,HB03,Ba15} 
in the perimetric coordinate system \cite{CJ37,Pe58}, 
a numerical method with the simplicity of a mesh calculation and the accuracy of a variational calculation. 
This method does not require analytical evaluations of integrals and computer times remain reasonable. 
It is found accurate in a variety of spectroscopic or collision applications \cite{Ba15}. 
In particular, in the case of antiprotonic helium, the accuracy on the energies matches the 
best available results in the literature and some other properties of the system 
can easily be computed \cite{BDS19}.  

The Lagrange-mesh method is briefly summarized in Sec.~\ref{LMM} 
and the conditions of the numerical calculations are determined. 
Energies and mean distances are presented in Sec.~\ref{emd}. 
These results are compared with the Born-Oppenheimer approximation and discussed in Sec.~\ref{disc}. 
Section \ref{conc} contains a summary and a conclusion. 
Atomic units are used throughout. 
\section{The Lagrange-mesh method}
\label{LMM}
\subsection{Summary}
\label{sum}
We study the quantal three-body system 
formed by a proton of mass $m_p =1836.152\,667\,5$, 
an antiproton of same mass $m_{\pb} = m_p$, 
and an electron of mass $m_e = 1$ in atomic units, 
interacting only through Coulomb forces. 
Fine structure and relativistic effects are not taken into account.  

The Schr\"odinger equation is solved in perimetric coordinates to avoid numerical problems 
with the singularities of the kinetic-energy operator and of the Coulomb interactions. 
The system of perimetric coordinates \cite{CJ37,Pe58} 
is defined by the three Euler angles $\psi, \theta, \phi$ and the three coordinates 
\beq
\left. \begin{array}{lll} 
x = r_{p\pb} + r_{p e} - r_{\pb e}, \\
y = r_{p\pb} - r_{p e} + r_{\pb e}, \\
z =-r_{p\pb} + r_{p e} + r_{\pb e},
\end{array} \right.
\eeqn{2.1}
involving the distances $r_{p\pb}$, $r_{p e}$, and $r_{\pb e}$ between the particles. 
The coordinates $x$, $y$ and $z$ vary over the $(0,\infty)$ interval. 
In perimetric coordinates, the Coulomb potential reads 
\beq
V(x,y,z) = -\frac{2}{x+y} - \frac{2}{x+z} + \frac{2}{y+z}. 
\eeqn{2.2}
The kinetic energy operator for $S$ states is given, e.g., in \Ref{Zh90}. 
The general expression for arbitrary states can be found in \Ref{HB03}. 

The wave function with total orbital momentum $L$ and natural parity $(-1)^L$ 
is expanded as \cite{HB03} 
\beq
\Psi^L_M = \sum_{K=0}^L {\cal D}_{MK}^L(\psi,\theta,\phi) \Phi_K^L(x,y,z). 
\eeqn{2.3}
where the ${\cal D}_{MK}^L (\psi,\theta,\phi)$ with $K \ge 0$ are parity-projected 
and normalized Wigner angular functions. 
In some cases, for $L > 0$, the sum can be truncated with excellent accuracy at some value $\Kmax$. 
For $\Kmax = 0$, the wave function presents a cylindrical symmetry along the $p\pb$ axis. 
The value of $\Kmax$ gives information about the departure from this symmetry. 

Let $u_i$, $v_j$, $w_k$ be the zeros of Laguerre polynomials of respective degrees $N_x$, $N_y$, $N_z$, 
and $h_x$, $h_y$, $h_z$ be three scale parameters with the dimension of a length in atomic units. 
The Lagrange-mesh method combines the three-dimensional mesh of $N_x N_y N_z$ points $(h_x u_i, h_y v_j, h_z w_k)$, 
a set of Lagrange functions $F^K_{ijk}(x,y,z)$ associated with each mesh point, 
and a Gauss quadrature consistent with this mesh \cite{BH86,Ba15,BDS19}.  
The Lagrange functions are constructed from Laguerre polynomials and their exponential weight function. 
They verify the Lagrange conditions 
\beq
F^K_{ijk}(h_x u_{i'}, h_y v_{j'}, h_z w_{k'}) \propto \delta_{ii'} \delta_{jj'} \delta_{kk'},
\eeqn{2.4}
i.e.\ each $F^K_{ijk}(x,y,z)$ vanishes at all mesh points except at the $ijk$ point. 
These functions are normed at the Gauss quadrature approximation which is used everywhere. 
The $\Phi_K^L(x,y,z)$ functions in \Eq{2.3} are expanded in the Lagrange basis as
\beq
\Phi_K^L(x,y,z) = \sum_{i=1}^{N_x} \sum_{j=1}^{N_y} \sum_{k=1}^{N_z} C_{Kijk}^L F^K_{ijk}(x,y,z).
\eeqn{2.5}
For each $L$ value, the coefficients are given by the mesh equations 
\beq
\sum_{Kijk} \Big\{ \langle F^{K'}_{i'j'k'}|T^L_{K'K}|F^K_{ijk} \rangle 
+ \left[ V(h_x u_i,h_y v_j,h_z w_k) - E_{L\nu} \right] 
\delta_{KK'} \delta_{ii'} \delta_{jj'} \delta_{kk'} \Big\} C_{Kijk}^L = 0,
\eeqn{2.6}
where $T^L_{K'K}$ is the matrix element of the kinetic-energy operator 
between functions ${\cal D}_{MK'}^L$ and ${\cal D}_{MK}^L$. 
The matrix elements of operator $T^L_{K'K}$ between Lagrange functions are 
computed with the Gauss quadrature associated with the mesh \cite{HB03}. 
The potential part in \Eq{2.6} is diagonal at this approximation because of the Lagrange property \rref{2.4} 
while the kinetic-energy part has a tridiagonal block structure with many zeros in the blocks. 
The matrix of this system is thus rather sparse. 
Its size is  $N_x N_y N_z (\Kmax +1)$. 
A limited number of eigenvalues of such big matrices can be obtained in rather short computing times 
with the code of \Ref{BN07}. 
The longest computations below with $\Kmax = 3$ take about half an hour on a fast workstation. 
By increasing order, the {\it physical} eigenvalues $E_{L\nu}$ are labeled with the quantum number $\nu$ 
starting from zero. 
Other eigenvalues are discarded as explained below in Sec.~\ref{cnc}. 
The eigenvectors corresponding to physical eigenvalues provide, from Eqs.~\rref{2.5} and \rref{2.3}, 
square-integrable approximations of the wave functions. 

This calculation is not variational for two reasons of different natures: 
(i) all eigenstates of $p \pb e$ are unbound and 
(ii) the Gauss quadrature approximation is not exact. 
Nevertheless, some eigenvalues and eigenvectors of this system may provide 
approximate but accurate energies and eigenfunctions of quasibound states or narrow resonances. 
The difficulty is to isolate these eigenvalues from those corresponding to 
square-integrable approximations of continuum states. 
The separation of eigenvalues with a physical meaning is easily performed 
by computing the mean distances between the particles.
For example, the mean distance between the proton and antiproton is simply given 
at the Gauss-quadrature approximation by 
\beq
\la r_{p \pb} \ra = \frac{1}{2} \sum_{Kijk} (C_{Kijk}^{L})^2 (h_x u_i + h_y v_j).
\eeqn{2.14}
Unphysical eigenvalues are indicated by very large electron-proton and electron-antiproton distances. 
Another consistent criterion is obtained from the probabilities 
\beq
P_L(K) = \sum_{ijk} (C_{Kijk}^L)^2. 
\eeqn{2.15}
One observes that physical states always have $P_L(0)$ close to unity 
in contrast with approximations of continuum states. 
Because many channels are open below the energies we are looking for, 
it is important to have an idea of the energy domain where to search. 
This is provided by the Born-Oppenheimer results of \Ref{Sa18}. 
They suggest the existence of quasibound states or narrow resonances 
in the domain $59 \leq L \leq 73$. 
\subsection{Conditions of the numerical calculations}
\label{cnc}
The painful part of the calculation is to determine near-optimal values 
of seven parameters: $h_x$, $h_y$, $h_z$, $N_x$, $N_y$, $N_z$, and $K$. 
Once these choices have been made, it becomes easy to reproduce our calculations. 

We have adopted the following strategy. 
First, since the system has a symmetry close to cylindrical, 
the Euler angles are chosen in such a way that the intrinsic axis 3 
is chosen along the $p\pb$ direction. 
With this convention, values of $K$ can be limited to a small $\Kmax$ for the high accuracies presented below. 
Here, results are obtained with $\Kmax = 3$ but there is no difference with $\Kmax = 2$ for $L \ge 68$. 
Second, the scale parameters $h_x$, $h_y$, $h_z$ are selected. 
To this end, for each $L$, a number of calculations are performed with relatively small matrices 
where these parameters are varied to find plateaus of stability, 
i.e., regions such that small variations of these parameters do not affect a number of stable digits 
of the lowest quasibound eigenvalue ($\nu = 0$). 
These fast preliminary calculations were performed with $N_x = N_y = N_z = 20$ and $\Kmax = 2$. 
As shown in Table \ref{tab:1} below, the parameters $h_x$, $h_y$, $h_z$ need not be known 
with a high accuracy. 
They depend on the total orbital momentum $L$. 
Third, the best possible accuracy is searched for by increasing $N_x$, $ N_y$, $N_z$. 
These values optimized on the second excited quasibound state ($\nu = 2$) are also 
given in Table \ref{tab:1}. 
Accurate results of the next excited states ($\nu = 3-5$) are obtained by increasing 
$N_y$ only, up to 38. 

The parameters are displayed in Table \ref{tab:1}. 
One observes that $h_x$ and $h_y$ monotonically increase with $L$. 
Parameter $h_z$ is significantly larger and presents a minimum near $L = 65$. 
While $N_x$ remains remarkably constant and rather small, 
$N_y$ and $N_z$ have contrasted evolutions. 
Above $L = 60$, $N_y$ increases and $N_z$ decreases. 
These tendencies are not valid below $L = 60$ because the smaller number of stable digits 
makes results poorly sensitive to the numbers of mesh points. 
Since the perimetric coordinates have no intuitive interpretation, 
the evolution of all these parameters is mainly phenomenological. 
\renewcommand{\baselinestretch}{0.6}
\begin{table}[ht]
\caption{Parameters of the Lagrange meshes.}
\begin{center}
\begin{tabular}{rcccccc}
\hline
 $L$ & $N_x$ & $N_y$ & $N_z$ & $h_x$ & $h_y$ & $h_z$ \\ 
\hline
 56 & 20 & 24 & 20 & 0.24 & 0.12 & 1.5 \\
 57 & 20 & 24 & 20 & 0.25 & 0.15 & 1.5 \\
 58 & 20 & 24 & 20 & 0.27 & 0.17 & 1.5 \\
 59 & 20 & 28 & 24 & 0.27 & 0.17 & 1.4 \\
 60 & 20 & 28 & 32 & 0.28 & 0.17 & 1.4 \\
 61 & 20 & 28 & 32 & 0.28 & 0.18 & 1.3 \\
 62 & 20 & 28 & 32 & 0.28 & 0.21 & 1.3 \\
 63 & 20 & 28 & 32 & 0.29 & 0.24 & 1.2 \\
 64 & 20 & 30 & 30 & 0.30 & 0.26 & 1.2 \\
 65 & 20 & 30 & 30 & 0.30 & 0.28 & 1.1 \\
 66 & 20 & 32 & 28 & 0.31 & 0.30 & 1.1 \\
 67 & 20 & 32 & 24 & 0.35 & 0.36 & 1.1 \\
 68 & 20 & 32 & 24 & 0.38 & 0.45 & 1.2 \\
 69 & 20 & 32 & 24 & 0.45 & 0.57 & 1.2 \\
 70 & 20 & 32 & 20 & 0.58 & 0.90 & 1.3 \\
 71 & 20 & 32 & 20 & 0.70 & 1.30 & 1.3 \\
 72 & 20 & 32 & 20 & 0.70 & 1.80 & 1.6 \\
\hline
\end{tabular}
\end{center}
\label{tab:1}
\end{table}
\renewcommand{\baselinestretch}{1.5}
\section{Energies and mean distances}
\label{emd}
Energies for $L = 56-72$ obtained from the three-body Schr\"odinger equation are presented in Table \ref{tab:2}. 
Their accuracies are based on the stability of digits when comparing calculations with ($N_x, N_y, N_z)$, 
$(N_x+2, N_y, N_z)$, $(N_x, N_y+2, N_z)$, and $(N_x, N_y, N_z+2)$ mesh points. 
The last digit may be uncertain within a few units. 
Unexpectedly, the easiest calculation concern $L = 70$ where the matrix size $N_x N_y N_z (\Kmax+1)$ is smaller 
and the physical energies are easy to find because they are not mixed with other eigenvalues. 
For $L$ smaller than about $60-62$, on the contrary, the physical eigenvalues are mixed with many non-physical ones 
and their localization may require more efforts. 

For each $L$ value, the ground quasibound state and a number of its vibrational excited states are shown. 
When stability of an excited eigenvalue could not be reached, the location is left empty in the table. 
The number of stable digits gives a rough indication of the size of the width. 
In the antiprotonic helium case, it was observed that $N$ stable decimal digits roughly correspond 
to a width $10^{-(N-1)}$ \cite{BDS19}. 
We expect that the same property holds here. 
Notice that, for eigenvalues with 14 stable digits, the width can even be smaller 
since this number of digits is limited by the computer accuracy. 

The widths assumed here correspond to the decay of the $p\pb e$ system 
into $p\pb + e$ (Auger electron emission or autodetachment) and $\pb$ + H (dissociation). 
According to \Ref{Sa18}, $\pb$ + H dissociation is the main separation process. 
This is confirmed by the mean values of the distances between particles shown below. 
Notice however that the main decay channel according to \Ref{Sa18} is due to spontaneous radiation, 
i.e., photon emission from the $pe$ subsystem to the hydrogen ground state 
leading to a dissociation into $\pb$ + H($1s$). 
Let us recall that the structure of the $pe$ subsystem is modelized as the lowest $n = 2$ Stark state 
in the Born-Oppenheimer study of \Ref{Sa18}. 
The widths are found to be a little smaller than the radiative width of the $2p$ state 
of a free hydrogen atom. 
This radiative channel is absent in the present study and thus not included in the assumed width. 
Since the wave functions are available in a rather simple form, this channel could be studied 
in the present approach but is delayed to an ulterior work including other electromagnetic transitions. 
\renewcommand{\baselinestretch}{0.6}
\begin{table}[t]
\caption{Energies $E_{L\nu}$ of the lowest vibrational states as a function of $L$ and $\nu$.}
\begin{center}
\begin{tabular}{rlll}
\hline
 $L$& $\nu = 0$           & $\nu = 1$         & $\nu = 2$         \\ 
\hline
 56 & $-0.15155         $ & $                 $ & $               $ \\
 57 & $-0.14774         $ & $-0.1453          $ & $               $ \\
 58 & $-0.144344        $ & $-0.14209         $ & $-0.1401        $ \\
 59 & $-0.14128113      $ & $-0.1393014       $ & $-0.13756       $ \\
 60 & $-0.1385388611    $ & $-0.13682153      $ & $-0.1353058     $ \\
 61 & $-0.13609647825   $ & $-0.13462534271   $ & $-0.13333407    $ \\
 62 & $-0.133936537967  $ & $-0.13269402621   $ & $-0.1316112327  $ \\
 63 & $-0.132043690689  $ & $-0.13101194704   $ & $-0.13012040959 $ \\
 64 & $-0.1304040491368 $ & $-0.129565256171  $ & $-0.12884774188 $ \\
 65 & $-0.1290048721099 $ & $-0.128341265648  $ & $-0.12778054460 $ \\
 66 & $-0.12783435033022$ & $-0.1273280835826 $ & $-0.126906785039$ \\
 67 & $-0.1268814293332 $ & $-0.1265142967803 $ & $-0.12621463030 $ \\
 68 & $-0.12613558871367$ & $-0.1258885427503 $ & $-0.12569184978 $ \\
 69 & $-0.12558636854493$ & $-0.12543866774143$ & $-0.12532474227 $ \\
 70 & $-0.12522187107221$ & $-0.12514963712438$ & $-0.125095923153$ \\
 71 & $-0.12502315145032$ & $-0.12499811020158$ & $-0.1249800756  $ \\
 72 & $-0.12494845876   $ & $-0.12494320      $ & $-0.1249392     $ \\
\hline
 $L$& $\nu = 3$           & $\nu = 4$         & $\nu = 5$         \\
 58 & $-0.1383          $ & $                 $ & $               $ \\
 59 & $-0.13599         $ & $-0.13464         $ & $               $ \\
 60 & $-0.1339698       $ & $-0.132798        $ & $-0.13176       $ \\
 61 & $-0.132203567     $ & $-0.13121623      $ & $-0.13035704    $ \\
 62 & $-0.1306702054    $ & $-0.129854541     $ & $-0.12914931    $ \\
 63 & $-0.1293523321    $ & $-0.128692500     $ & $-0.128127166   $ \\
 64 & $-0.1282359709    $ & $-0.1277159374    $ & $-0.127275102   $ \\
 65 & $-0.1273084143    $ & $-0.126912127     $ & $-0.126580404   $ \\
 66 & $-0.1265574681    $ & $-0.1262687282    $ & $-0.12603061    $ \\
 67 & $-0.1259708842    $ & $-0.1257731515    $ & $-0.125613008   $ \\
 68 & $-0.12553565717   $ & $-0.1254118104    $ & $-0.125313632   $ \\
 69 & $-0.125236885915  $ & $-0.1251619068    $ & $-0.12511661    $ \\
 70 & $-0.1250558121336 $ & $-0.1250257350    $ & $-0.12500309    $ \\
\hline
\end{tabular}
\end{center}
\label{tab:2}
\end{table}
\renewcommand{\baselinestretch}{1.5}

From $L = 56$ to 59, the number of stable digits is small. 
The widths are expected to be large. 
No or few excited states could be obtained. 
They should be resonances too broad for the present static calculations. 
From $L = 60$ to 70, we display five vibrational excited states 
(see Sec.~\ref{disc} for a justification of this interpretation). 
The states become narrower when $L$ increases and broader when $\nu$ increases. 
For $L = 70$, the $\nu = 5$ energy is very close to the $\pb$ + H($n = 2$) dissociation threshold. 
This excited vibrational state is probably the highest to be quasibound. 
For $L = 71$, only the $\nu = 0$ state is quasibound. 
We display the energies of the first two resonances which are extremely narrow. 
No quasibound state is found for $L = 72$ but the ground-state resonance 
is also very narrow.  
 
\begin{figure}[ht]
\setlength{\unitlength}{1 mm}
\begin{picture}(160,60) (0,115) 
\put(0,0){\includegraphics[width=1.0\textwidth]{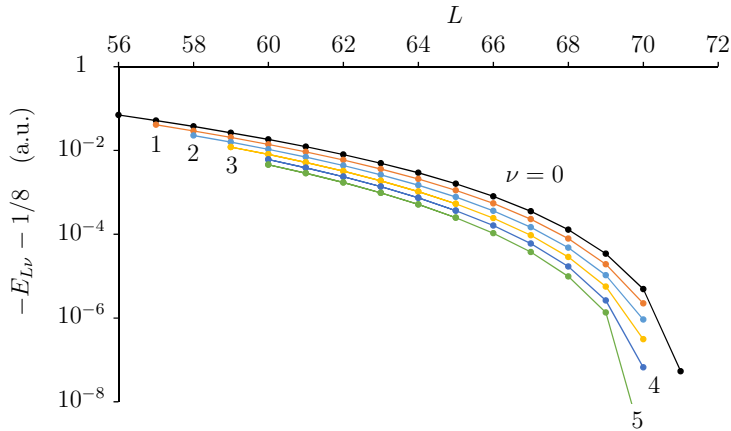}}
\end{picture} \\
\caption{Binding energies $-E_{L\nu}-1/8$ with respect to the $\pb$ + H($n = 2$) threshold as a function of $L$. 
The curves are labeled by the vibrational quantum number $\nu$. }
\label{fig:e}
\end{figure}

The binding energies $-E_{L\nu}-1/8$ with respect to the $\pb$ + H($n = 2$) dissociation threshold 
are displayed in logarithmic scale in Fig.~\ref{fig:e}. 
One observes that they decrease monotonically, faster than exponentially, as a function of $L$. 
For each $L$ value except for $L \geq 70$, the points in the figure present an almost constant spacing. 
This corresponds to exponentially decreasing binding energies as a function of $\nu$ 
in the considered range $\nu \leq 5$. 
\renewcommand{\baselinestretch}{0.6}
\begin{table}[t]
\caption{Mean distances between the particles as a function of $L$ and $\nu$.}
\begin{center}
\begin{tabular}{rrlll}
\hline
 $L$ & $\nu$ & $\la r_{p \pb} \ra$ & $\la r_{p e} \ra$ & $\la r_{\pb e} \ra$ \\ 
\hline
 56 & 0 & 3.96      & 12.5      & 15.4       \\
 57 & 0 & 4.23      & 10.6      & 13.8       \\
    & 1 & 4.5       &           &            \\
 58 & 0 & 4.504     & 9.6       & 13.0       \\
    & 1 & 4.7       &           &            \\
 59 & 0 & 4.80035   & 8.985     & 12.677     \\
    & 1 & 5.208     & 8.82      & 12.88      \\
    & 2 & 5.6       &           &            \\
 60 & 0 & 5.12423   & 8.5280    & 12.5026    \\
    & 1 & 5.5792    & 8.266     & 12.650     \\
    & 2 & 6.06      & 8.1       & 13.0       \\
 61 & 0 & 5.4865215 & 8.127147  & 12.421703  \\
    & 1 & 5.993173  & 7.87762   & 12.62937   \\
    & 2 & 6.5430    & 7.6553    & 12.9061    \\
 62 & 0 & 5.897680  & 7.77372   & 12.43425   \\
    & 1 & 6.465661  & 7.54210   & 12.71805   \\
    & 2 & 7.085388  & 7.32942   & 13.07158   \\
 63 & 0 & 6.3720818 & 7.4546917 & 12.5407382 \\
    & 1 & 7.0153989 & 7.2428201 & 12.9163477 \\
    & 2 & 7.7208651 & 7.049496  & 13.3718622 \\
 64 & 0 & 6.9301622 & 7.1637094 & 12.7541037 \\
    & 1 & 7.6689101 & 6.9717212 & 13.2414497 \\
    & 2 & 8.4837312 & 6.7978713 & 13.8224752 \\
 65 & 0 & 7.6027009 & 6.8961919 & 13.0991423 \\
    & 1 & 8.4665215 & 6.7237866 & 13.7271771 \\
    & 2 & 9.425775  & 6.5691080 & 14.468225  \\
 66 & 0 & 8.4389209 & 6.6486442 & 13.6196868 \\
    & 1 & 9.4736262 & 6.4955777 & 14.4335341 \\
    & 2 & 10.63183  & 6.359864  & 15.388694  \\
 67 & 0 & 9.5236837 & 6.4183166 & 14.3951181 \\
    & 1 & 10.804703 & 6.2846998 & 15.4698911 \\
    & 2 & 12.251974 & 6.1681192 & 16.729417  \\
 68 & 0 & 11.018548 & 6.2031032 & 15.5807073 \\
    & 1 & 12.680887 & 6.0897259 & 17.0520249 \\
    & 2 & 14.578629 & 5.9930624 & 18.778600  \\
 69 & 0 & 13.275801 & 6.0018471 & 17.5208472 \\
    & 1 & 15.588816 & 5.9106846 & 19.6608992 \\
    & 2 & 18.25717  & 5.8355943 & 22.17830   \\
 70 & 0 & 17.221213 & 5.8160442 & 21.1321144 \\
    & 1 & 20.798740 & 5.7508863 & 24.5630015 \\
    & 2 & 24.959507 & 5.6997582 & 28.601044  \\
 71 & 0 & 25.884281 & 5.6568797 & 29.4453979 \\
    & 1 & 32.344848 & 5.6214789 & 35.802946  \\
    & 2 & 39.9390   & 5.594804  & 43.3144    \\
 72 & 0 & 50.412    & 5.554296  & 53.669     \\
\hline
\end{tabular}		    
\end{center}
\label{tab:3}
\end{table}
\renewcommand{\baselinestretch}{1.5}
In Table \ref{tab:3} are presented mean values of the distances $\la r_{p \pb} \ra$ between proton and antiproton, 
$\la r_{p e} \ra$ between proton and electron, and $\la r_{\pb e} \ra$ between antiproton and electron. 
Here also, only stable digits are kept except the last one which may be uncertain within a few units. 
The number of presented decimal digits is limited to a maximum of seven. 
In all cases, the accuracy does not exceed 9 decimal places in the range $65-70$. 
Indeed, since the method is approximately variational, an error $\epsilon$ on wave functions 
and mean values corresponds to an error of about $\epsilon^2$ on energies. 
A $10^{-9}$ accuracy on the distances means that the accuracy of the energies in this range 
could be better than $10^{-14}$. 

\begin{figure}[ht]
\setlength{\unitlength}{1 mm}
\begin{picture}(160,60) (0,115) 
\put(0,0){\includegraphics[width=1.0\textwidth]{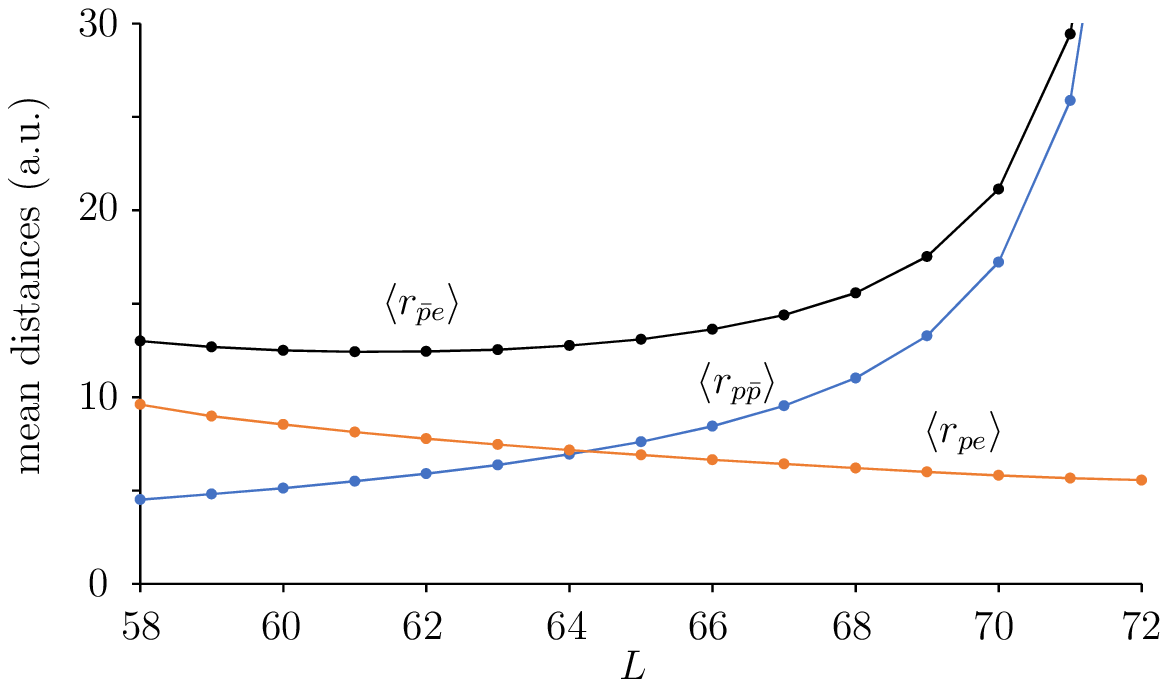}}
\end{picture} \\
\caption{Mean distances $\la r_{p \pb} \ra$, $\la r_{p e} \ra$, 
and $\la r_{\pb e} \ra$ between the particles for $\nu = 0$.}
\label{fig:dist}
\end{figure}

Let us first consider $\la r_{p e} \ra$. 
This mean distance decreases monotonically from the large value 12.5 at $L = 56$ 
to 5.65 at $L = 71$ (see also Fig.~\ref{fig:dist}). 
It decreases when $\nu$ increases. 
Around $L = 69$, $\la r_{p e} \ra$ is very close to 6, i.e., the mean distance between 
the proton and electron of a free hydrogen atom in an $n = 2$ state. 
This nicely corresponds to the assumption in the model of \Ref{Sa18}. 
For lower $L$ values, higher excitations of hydrogen also play a role. 
Beyond $L = 69$ and for some higher vibrational states, a significant component 
of hydrogen in its ground state must also be present. 

Let us now turn to the proton-antiproton mean distance $\la r_{p \pb} \ra$. 
It monotonically increases with both $L$ and $\nu$. 
The increase with $L$ follows from the centrifugal effect. 
The increase with $\nu$ corresponds to a broadening due to vibrational excitation. 
This is discussed in another way in Sec.~\ref{disc}. 

The antiproton-electron mean distance $\la r_{\pb e} \ra$ is always 
larger than the proton-electron one $\la r_{p e} \ra$ as  expected 
from the electrostatic repulsion between these particles. 
It also increases with the vibrational quantum number $\nu$. 

\begin{figure}[ht]
\setlength{\unitlength}{1 mm}
\begin{picture}(160,60) (-5,115) 
\put(0,0){\includegraphics[width=1.0\textwidth]{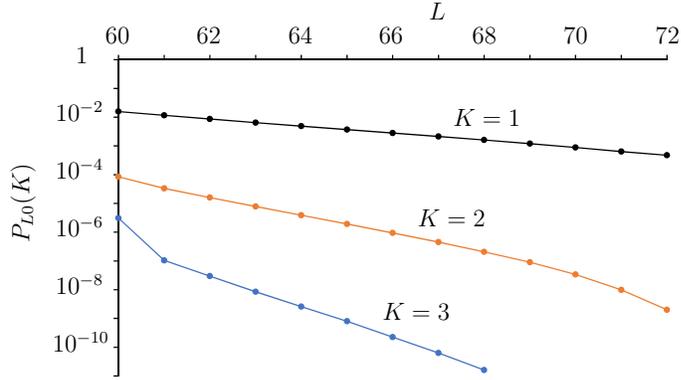}}
\end{picture} \\
\caption{Probabilities of the $K > 0$ components as a function of $L$ for $\nu = 0$.}
\label{fig:prob}
\end{figure}

As mentioned before, the probabilities of $K$ components give indications 
about the cylindrical symmetry of the system. 
The $K > 0$ probabilities are displayed with a maximum of ten decimal digits 
in Table \ref{tab:4} and illustrated by Fig.~\ref{fig:prob}. 
They are quite inaccurate for $L < 60$. 
For $L > 65$, an excellent accuracy on the energies is already obtained with $\Kmax = 2$, 
or even $\Kmax = 1$ for $L \geq 70$. 
On the contrary, for $L < 60$, values larger then 3 could be necessary 
in more elaborate calculations of resonance properties. 

\renewcommand{\baselinestretch}{0.6}
\begin{table}[t]
\caption{Probabilities $P_{L\nu}(K)$ of the $K > 0$ components as a function of $L$ and $\nu$.}
\begin{center}
\begin{tabular}{rrlll}
\hline
 $L$ & $\nu$ & $P_{L\nu}(K=1)$ & $P_{L\nu}(K=2)$ & $P_{L\nu}(K=3)$ \\ 
\hline
 56 & 0 & 0.082        &              &              \\
 57 & 0 & 0.047        & 0.003        &              \\
 58 & 0 & 0.031        & 0.0006       &              \\
 59 & 0 & 0.02162      & 0.000179     & 0.000003     \\
    & 1 & 0.024        & 0.0008       &              \\
 60 & 0 & 0.015677     & 0.0000841    & 0.0000031    \\
    & 1 & 0.01551      & 0.000136     & 0.000003     \\
    & 2 & 0.0169       & 0.0009       &              \\
 61 & 0 & 0.011536186  & 0.000033516  & 0.000000104  \\
    & 1 & 0.0110244    & 0.0000512    & 0.00000034   \\
    & 2 & 0.01050      & 0.00010      & 0.000010     \\
 62 & 0 & 0.0085988    & 0.0000161    & 0.00000003   \\
    & 1 & 0.0080172    & 0.00002199   & 0.00000007   \\
    & 2 & 0.0074112    & 0.00002714   & 0.00000013   \\
 63 & 0 & 0.0064633290 & 0.0000078829 & 0.0000000085 \\
    & 1 & 0.0059116281 & 0.0000098772 & 0.0000000174 \\
    & 2 & 0.00536622   & 0.000011321  & 0.0000000274 \\
 64 & 0 & 0.0048875929 & 0.0000038902 & 0.0000000026 \\
    & 1 & 0.004401258  & 0.0000045151 & 0.0000000047 \\
    & 2 & 0.003939311  & 0.0000048635 & 0.0000000067 \\
 65 & 0 & 0.0037101822 & 0.0000019199 & 0.0000000008 \\
    & 1 & 0.0032987199 & 0.0000020724 & 0.0000000013 \\
    & 2 & 0.002920324  & 0.0000021061 & 0.0000000017 \\
 66 & 0 & 0.0028202995 & 0.0000009394 & 0.0000000002 \\
    & 1 & 0.002482245  & 0.0000009447 & 0.0000000003 \\
    & 2 & 0.002179932  & 0.0000009064 & 0.0000000004 \\
 67 & 0 & 0.0021402927 & 0.0000004505 & 0.0000000001 \\
    & 1 & 0.0018698693 & 0.0000004224 & 0.0000000001 \\
    & 2 & 0.0016343534 & 0.0000003824 & 0.0000000001 \\
 68 & 0 & 0.0016148758 & 0.0000002080 & 0.0000000000 \\
    & 1 & 0.0014052474 & 0.0000001819 & 0.0000000000 \\
    & 2 & 0.0012276417 & 0.0000001553 & 0.0000000000 \\
 69 & 0 & 0.0012044099 & 0.0000000895 & 0.0000000000 \\
    & 1 & 0.0010494638 & 0.0000000732 & 0.0000000000 \\
    & 2 & 0.000922250  & 0.0000000590 & 0.0000000000 \\
 70 & 0 & 0.0008816009 & 0.0000000338 & 0.0000000000 \\
    & 1 & 0.0007769888 & 0.0000000260 & 0.0000000000 \\
    & 2 & 0.000694041  & 0.0000000199 & 0.0000000000 \\
 71 & 0 & 0.0006344935 & 0.0000000099 & 0.0000000000 \\
    & 1 & 0.0005759509 & 0.0000000073 & 0.0000000000 \\
    & 2 & 0.0005303019 & 0.0000000054 & 0.0000000000 \\
 72 & 0 & 0.000474112  & 0.0000000020 & 0.0000000000 \\
\hline
\end{tabular}		    
\end{center}
\label{tab:4}
\end{table}
\renewcommand{\baselinestretch}{1.5}
Table \ref{tab:4} shows contrasting behaviors for the probabilities $P_{L\nu}(K)$ 
of vibrational excited states. 
For $K = 1$, the probabilities decrease when $\nu$ increases. 
For $K = 2$, they increase for $L < 66$ and decrease for $L > 66$. 
For $K = 3$, they always increase with $\nu$. 
The dependence of $P_{L0}(K)$ on $L$ is illustrated in Fig.~\ref{fig:prob}. 
All probablities decrease monotonically with $L$. 
The $K = 1$ probabilities are the largest. 
The $K = 2$ probabilities are smaller than $10^{-4}$. 
The $K = 3$ probabilities become smaller than $10^{-11}$ beyond $L = 68$. 
They explain why this component does affect the energies in this range. 
The smallness of the $K > 0$ probabilities implies that $P_{L\nu}(0)$ is always close to unity. 
Its smallest value, near 0.92, is for $L = 56$. 
\clearpage
\section{Discussion}
\label{disc}
\begin{figure}[ht]
\setlength{\unitlength}{1 mm}
\begin{picture}(160,55) (-5,110) 
\put(0,0){\includegraphics[width=1.0\textwidth]{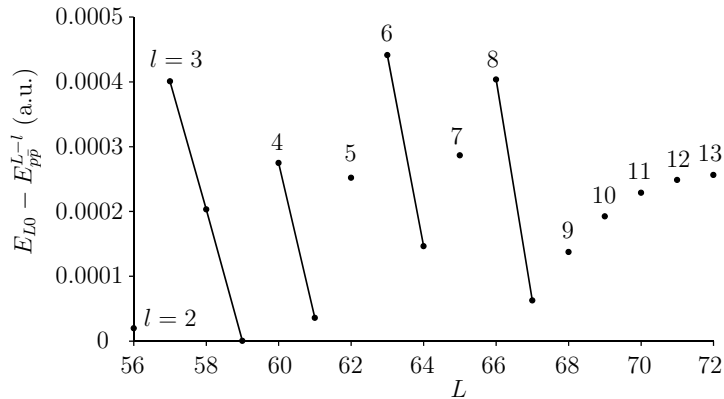}}
\end{picture} \\
\caption{Electron emission energy $E_{L0} - E_{p\pb}^{L-l}$ 
for the lowest electron orbital momentum $l$ as a function of $L$, 
where $E_{p\pb}^{L-l}$ is the energy of the highest open threshold below $E_{L0}$.}
\label{fig:seuils}
\end{figure}

The electron emission energies $E_{L0} - E_{p\pb}^{L-l}$, 
where $E_{p\pb}^{L-l}$ is the highest $p\pb$ threshold energy below $E_{L0}$ 
corresponding to the lowest orbital momentum $l$ of the emitted electron, 
are displayed in Fig.~\ref{fig:seuils} as a function of $L$. 
The points are labeled with the value of this minimal orbital momentum $l$ 
(denoted as $l_0$ in Table II of \Ref{Sa18}). 
One observes that this energy is very small at $L = 59$, 
which may explain why the search for optimal parameters was very difficult in that case. 
It is also quite small for $L = 56$ and 61. 
These energies tend to more constant values at high $L$, 
including for the $L = 72$ narrow resonance. 

The states from $L = 60$ to 71 have long lifetimes with respect to the channels $p \pb + e$ 
because this Auger electron emission (or autodetachment) is a slow process. 
The spontaneous emission of an electron is strongly hindered in these states 
because the electron can only be emitted with a rather high orbital momentum $l \geq 3$ \cite{Ru70,Sa18}. 
The same effect explains the long lifetimes of some levels of antiprotonic helium (\Ref{BDS19} and references therein). 
The large number of stable digits indicates that all couplings with continuum states are weak 
in this range of $L$ values in agreement with the findings of \Ref{Sa18}. 
This is also confirmed by the absence or small number of non-physical eigenvalues 
among the physical energies in this domain. 

The quantum number $\nu$ is interpreted here as representing a vibrational excitation. 
In order to deepen this interpretation, we now perform a comparison with the Born-Oppenheimer picture 
as used in \Ref{Sa18}. 
To this end, we solve the Born-Oppenheimer equation 
\beq
\left( - \frac{1}{2} \Delta - \frac{1}{r_{pe}} + \frac{1}{r_{\pb e}} - \frac{1}{R} \right) \chi_{nm}(\ve{r}_{pe},R) 
= \mathcal{E}_{nm}(R) \chi_{nm}(\ve{r}_{pe},R),
\eeqn{4.1}
where $R \equiv r_{p \pb}$ is fixed and the Laplacian $\Delta$ corresponds to the proton-electron coordinate $\ve{r}_{p e}$. 
The energies depend on the parameter $R$ and on the magnetic quantum number $m$. 
They are labeled by the excitation quantum number $n$ starting from 0. 
This equation is separable in confocal elliptic coordinates (or prolate spheroidal coordinates) \cite{WHM60}. 
Here, we solve it for $m = 0$ with the Lagrange-mesh method as explained in \Ref{VB06}. 

For the coordinate $\xi = (r_{pe} + r_{\pb e})/R - 1$ defined in the interval $(0,\infty)$, 
a Lagrange-Laguerre mesh with $N_\xi$ points and a scale parameter $h$ is employed. 
For the coordinate $\eta = (r_{pe} - r_{\pb e})/R$ defined in the interval $(-1,1)$, 
a Lagrange-Legendre mesh with $N_\eta$ points is used. 
For $m = 0$, the azimutal angle $\varphi$ of $\ve{r}_{pe}$ does not play a role. 
Since the calculation is very fast, a rough optimization is sufficient. 
We use $N_\xi = 30$ and $N_\eta = 20$. 
The scale parameter is given by $h = \max(6/R-1,2/R)$. 
The lowest energies $\mathcal{E}_{00}(R)$ correspond to those labeled (0,0,0) in Refs.~\cite{WHM60,Sa18}. 
The first excited ones $\mathcal{E}_{10}(R)$ correspond to those labeled (1,0,0) in these references. 
These energies are obtained with at least nine stable decimal digits. 
When rounded at the fifth decimal digit, they perfectly agree with the results of \Ref{WHM60}. 

\begin{table}[hbt]
\caption{Born-Oppenheimer energies $E^{BO}_{L\nu}$ and mean distances $\la R \ra_{L\nu}$ 
of the lowest vibrational states as a function of $L$.}
\begin{center}
\begin{tabular}{rcccccc}
\hline
 $L$ & $E^{BO}_{L0}$ & $\la R \ra_{L0}$ & $E^{BO}_{L1}$ & $\la R \ra_{L1}$ 
		 & $E^{BO}_{L2}$ & $\la R \ra_{L2}$ \\
\hline
  58 & $-0.1439567$ &  4.5952 & $-0.1417104$ &  4.9849 & $-0.1397098$ &  5.4040 \\
  59 & $-0.1409907$ &  4.8804 & $-0.1390290$ &  5.3055 & $-0.1372899$ &  5.7638 \\
  60 & $-0.1383227$ &  5.1987 & $-0.1366258$ &  5.6654 & $-0.1351293$ &  6.1703 \\
  61 & $-0.1359388$ &  5.5575 & $-0.1344874$ &  6.0739 & $-0.1332150$ &  6.6349 \\
  62 & $-0.1338256$ &  5.9669 & $-0.1326009$ &  6.5440 & $-0.1315350$ &  7.1736 \\
  63 & $-0.1319704$ &  6.4413 & $-0.1309544$ &  7.0937 & $-0.1300774$ &  7.8091 \\
  64 & $-0.1303614$ &  7.0012 & $-0.1295362$ &  7.7496 & $-0.1288310$ &  8.5753 \\
  65 & $-0.1289873$ &  7.6779 & $-0.1283352$ &  8.5528 & $-0.1277847$ &  9.5246 \\
  66 & $-0.1278375$ &  8.5217 & $-0.1273408$ &  9.5701 & $-0.1269278$ & 10.7438 \\
  67 & $-0.1269017$ &  9.6198 & $-0.1265423$ & 10.9188 & $-0.1262493$ & 12.3869 \\
  68 & $-0.1261701$ & 11.1386 & $-0.1259291$ & 12.8271 & $-0.1257375$ & 14.7552 \\
  69 & $-0.1256327$ & 13.4434 & $-0.1254894$ & 15.7992 & $-0.1253790$ & 18.5175 \\
  70 & $-0.1252778$ & 17.5009 & $-0.1252083$ & 21.1596 & $-0.1251568$ & 25.4150 \\
  71 & $-0.1250861$ & 26.4883 & $-0.1250624$ & 33.1343 & $-0.1250453$ & 40.9531 \\
\hline										   
\end{tabular}
\end{center}
\label{tab:5}
\end{table}
The energies $\mathcal{E}_{10}(R)$ are then used as a proton-antiproton potential 
in radial Schr\"odinger equations with reduced mass $m_p/2$. 
These equations are solved for $L = 58$ to 71 with the Numerov algorithm. 
Above $R = 2$, the steps 0.1, 0.05, and 0.02 have been used up to $R = 100$. 
Below $R = 2$, these high $L$ effective potentials are approximated only by the centrifugal barriers. 
The results are displayed in Table \ref{tab:5} for $h = 0.02$. 
The energies $E^{BO}_{L\nu}$ of the three calculations agree within at least 7 decimal digits for the three lowest vibrational states. 
This accuracy is better than the difference between the present energies and the energies $E_{BO}$ of the first column 
of Table II in \Ref{Sa18} which are based on interpolations of the results of \Ref{WHM60}. 
For the mean proton-antiproton distances, an absolute accuracy better than about $10^{-5}$ is obtained with $h = 0.02$ 
for $\nu = 0$ and a little less good for $\nu = 1$ and 2. 
The accuracy is poorer for $L = 71$. 
For this $L$ value, the existence of vibrational excited states	below the $\pb$ + H($n = 2$) threshold 
in the Born-Oppenheimer approach is in contradiction with the three-body results. 

\begin{figure}[ht]
\setlength{\unitlength}{1 mm}
\begin{picture}(160,55) (-5,105) 
\put(0,0){\includegraphics[width=0.9\textwidth]{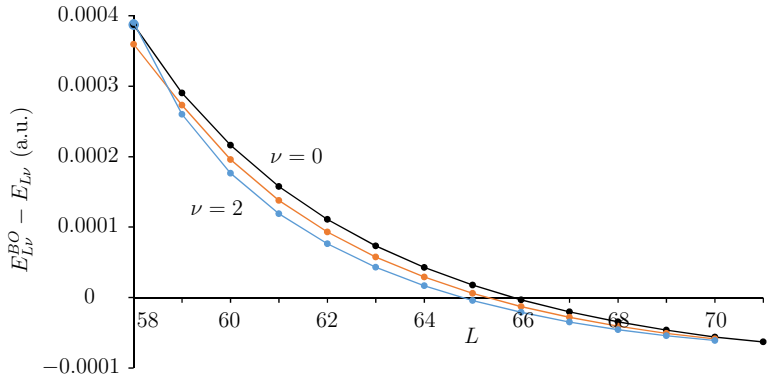}}
\end{picture} \\
\caption{Differences of the energies $E^{BO}_{L\nu}$ of Table \ref{tab:5} computed at the Born-Oppenheimer approximation 
and the three-body energies $E_{L\nu}$ of Table \ref{tab:2} for $L = 58-71$ and $\nu = 0-2$.}
\label{fig:EBO}
\end{figure}

In Fig.~\ref{fig:EBO}, we display for $\nu = 0-2$ the differences of the energies $E^{BO}_{L\nu}$ of Table \ref{tab:5} 
computed with the Born-Oppenheimer approach and the quantal three-body energies $E_{L\nu}$ of Table \ref{tab:2}. 
These differences monotonically decrease. 
They are positive below $L = 65-66$ and negative above. 
They also decrease with increasing $\nu$. 
A figure involving the improved Born-Oppenheimer energies of the second column of Table II in \Ref{Sa18} 
would be hardly different at the present scale. 
The error of the Born-Oppenheimer approximation is always smaller than $4 \times 10^{-4}$ atomic units. 
This good agreement confirms the interpretation of $\nu$ as a vibrational quantum number. 

This picture is also confirmed by the ratios of mean distances $\la R \ra_{L\nu}$ of Table \ref{tab:5} 
obtained at the Born-Oppenheimer approximation and the quantal mean distances $\la r_{p \pb} \ra_{L\nu}$ of Table \ref{tab:3}. 
With the exception of $L = 71$ for which only one quantal state is below the $\pb$ + H($n = 2$) threshold, 
the Born-Oppenheimer mean distances are between one and two percent larger than the quantal ones. 
This is true for the three presented $\nu$ values. 
For $L = 71$, the $\nu = 0$ ratio is smaller than unity, i.e., 0.9815.
\begin{figure}[ht]
\setlength{\unitlength}{1 mm}
\begin{picture}(160,55) (0,105) 
\put(0,0){\includegraphics[width=0.9\textwidth]{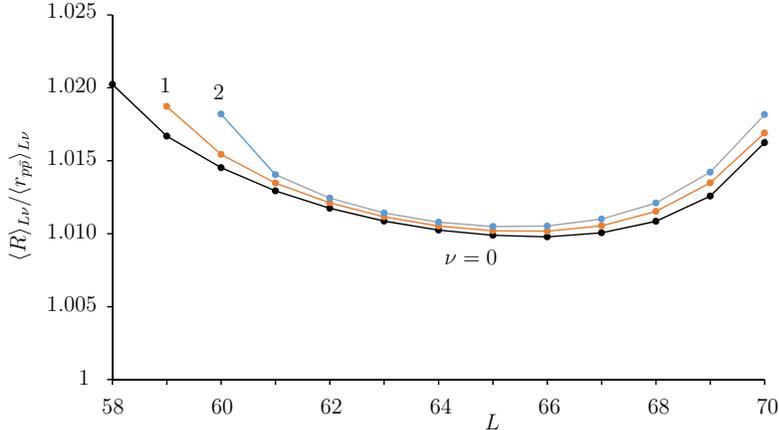}}
\end{picture} \\
\caption{Ratios of mean distances $\la R \ra_{L\nu}$ of Table \ref{tab:5} computed at the Born-Oppenheimer approximation 
to the corresponding mean distances $\la r_{p \pb} \ra_{L\nu}$ of Table \ref{tab:3} for $L = 58-70$ and $\nu = 0-2$.}
\label{fig:RBO}
\end{figure}

Finally, let us mention that we found some evidence around $L = 65$ for other resonances with fewer stable digits 
and with mean proton-electron distances close to 3. 
These resonances would correspond to configurations with a hydrogen atom mainly in its ground state. 
Such structures would contradict the Born-Oppenheimer picture of possible structures discussed in \Ref{Sa18}. 
Unfortunately, since these resonances should be broader, the present approach is not adequate for confirming 
their existence and evaluating their properties. 
\section{Conclusion}
\label{conc}
With an accurate three-body, approximately variational, calculation on a Lagrange mesh, 
we establish the existence of quasibound states below the $\pb$ + H($n = 2$) threshold 
with extremely small widths for spontaneous dissociation. 
The existence of these states was predicted in the Born-Oppenheimer study of \Ref{Sa18}. 
We show that several narrow vibrational states exist over a range of $L$ values. 
We have studied the six lowest vibrational states between $L = 60$ and 70 and the only existing such state at $L = 71$. 
Below $L = 60$, rather narrow resonances still exist but are less well described by the present static study. 
In all cases, the corresponding configurations are not far from a cylindrical symmetry around the proton-antiproton axis. 

This vibrational interpretation is confirmed by a Born-Oppenheimer calculation similar to the one 
performed in \Ref{Sa18}. 
The accuracy of this approximation in the present case is thus determined. 
It is found to improve with increasing $L$, reaching about $10^{-4}$ atomic units for the energies. 

Accurate values are also obtained for the three mean distances between the particles. 
The mean distances between the proton and electron indicate that the hydrogen-atom-like substructure 
in the three-body system is in a superposition of excited states, 
sometimes dominated by the $n = 2$ state assumed in \Ref{Sa18}. 
The proton-antiproton mean distances increase as expected from the centrifugal effect. 
They are well described by the Born-Oppenheimer approximation with an overestimation between 1 and 2 percents. 

The number of stable digits obtained for the energies gives indications about the widths 
of the quasibound states which are found very small between $L = 60$ and 71. 
These widths only correspond to spontaneous dissociations into $p\pb + e$ or $\pb$ + H.  
In fact, according to \Ref{Sa18}, the main decay channel of these states is radiative: 
emission of a photon with dissociation into $\pb$ + H($1s$). 
This information is based on a Born-Oppenheimer approach with the restrictive assumption 
that the hydrogen-atom subsystem is in a specific $n = 2$ state. 
This assumption is only partly in agreement with our results. 
A quantal three-body study of radiative transitions from quasibound to unbound states 
leading to this dissociation as well as of the electromagnetic deexcitations 
of the vibrational excited states will be the object of a future work. 
\end{document}